\documentclass[twocolumn]{aastex631}
\usepackage{subfigure}

\shorttitle{Orbital-period Changes of LMXBs Driven by Magnetic Braking}
\shortauthors{Fan et al.}

\begin{document}


\title{Orbital-period Changes of Low-mass X-ray Binaries Driven by Magnetic Braking}


\author[0009-0006-0116-5175]{Yun-Ning Fan}
\affil{School of Science, Qingdao University of Technology, Qingdao 266525, People’s Republic of China;chenwc@pku.edu.cn}
\author[0000-0003-2506-6906]{Yong Shao}
\affil{Department of Astronomy, Nanjing University, Nanjing 210023, People’s Republic of China}
\affil{Key Laboratory of Modern Astronomy and Astrophysics (Nanjing University), Ministry of Education, Nanjing 210023, People’s Republic of China}
\author[0000-0002-0785-5349]{Wen-Cong Chen}
\affil{School of Science, Qingdao University of Technology, Qingdao 266525, People’s Republic of China;chenwc@pku.edu.cn}
\affil{School of Physics and Electrical Information, Shangqiu Normal University, Shangqiu 476000, People’s Republic of China}



\begin{abstract}
The magnetic braking (MB) plays an important role in driving the evolution of low-mass X-ray binaries (LMXBs). The modified MB prescription, convection and rotation boosted (CARB) model, is very successful in reproducing the detected mass-transfer rates of persistent neutron star (NS) LMXBs. In this work, we investigate whether the CARB MB prescription could account for the formation and evolution of some NS and black hole (BH) LMXBs with an observed orbital period derivative. Using the MESA code, we perform a detailed binary evolution model for six NS and three BH LMXBs. Our simulations find that the CARB MB prescription can successfully reproduce the observed donor-star masses, orbital periods, and period derivatives of four NS LMXBs and one BH LMXB. Our calculated effective temperatures are in good agreement with the detected spectral types of two NS LMXBs and one BH LMXB. However, the standard MB model is difficult to produce the observed period derivatives of those LMXBs experiencing a rapid orbital shrinkage or expansion.

\end{abstract}


\section{Introduction}
After the first Galactic X-ray source was discovered \citep[Sco X-1,][]{Giac62}, those strong Galactic X-ray sources were proposed to be accreting neutron stars (NSs) or black holes (BHs) in X-ray binaries \citep{Taur06}. Their typical X-ray luminosities are in the range from $10^{34}$ to $10^{38}~\rm erg\,s^{-1}$ (the luminosities of ultraluminous X-ray sources can exceed $10^{39}~\rm erg\,s^{-1}$), which is $1- 5$ orders of magnitude higher than the total luminosity of the Sun. Over 90\% of the strong Galactic X-ray sources can be classified into the following two populations: high-mass X-ray binaries (HMXBs) and low-mass X-ray binaries (LMXBs), with a donor-star mass of $\ge 10~M_{\odot}$ and  $\le 1-2~M_{\odot}$, respectively. As an old stellar population (with an age $(5-13)\times 10^{9}~ \rm yr$), about 200 LMXBs have been discovered in the Galaxy \citep{Liu07}, with a concentrated distribution towards Galactic center, and a widespread around Galactic plane \citep{van95}. Their formation is still puzzling at present \citep{Taur06,Li15}. 

In an NS LMXB, the NS has accreted the material and angular momentum from its donor star via Roche lobe overflow and evolves toward a millisecond pulsar \citep{alpa82}. After the H-rich of the donor star is stripped, the LMXB evolves into a binary pulsar with a He/CO white dwarf (WD). Subsequently, the He/CO WD in a compact binary pulsar fills its Roche lobe due to an orbital decay through gravitational radiation (GR) and initiates to transfer mass onto the NS, driving the system to evolve into an ultracompact X-ray binary that are the potential low-frequency gravitational wave (GW) sources in the Galaxy \citep{taur18,chen20b}. Therefore, the formation and evolution of LMXBs are very significant in understanding the stellar and binary evolution theory, the origin of binary millisecond pulsars, and low-frequency GW sources.

In LMXBs with a short orbital period ($\le 12~\rm hr$), the progenitor radius of the NS must be much larger than the current orbital separation \citep{Taur06}. Therefore, a common-envelope stage or a large angular momentum loss (AML) is inevitable during their evolution.
As the primary mechanism for AML, the GR dominates the orbital evolution and mass transfer of those LMXBs with orbital periods shorter than $2-3$ hours \citep{Weis16}.  On the contrary, magnetic braking (MB) plays an important role in driving the mass transfer of LMXBs with an orbital period longer than 3 hours \citep{rapp83}. To account for the formation of BH LMXBs, it requires invoking other AML mechanisms such as anomalous MB of Ap/Bp stars \citep{Just06}, a surrounding circumbinary disk \citep{Chen15}, as well as the dynamical friction of dark matter \citep{qin24b}.

In LMXBs including a main-sequence (MS) donor star with a convective envelope and radiative core, the MB plays a vital role in influencing their evolution. Adopting the standard MB prescription \citep{rapp83}, the calculated mass-transfer rates are at least an order of magnitude lower than the observed values in some LMXBs with short orbital periods \citep{Pods02}. To interpret the detected highly mass-transfer rate of Sco X-1, \cite{pavl16} proposed a wind-boosted MB prescription that the wind loss in evolved stars is much stronger than that in MS stars. \cite{chen17} thought that Sco X-1 could have evolved from an intermediate-mass X-ray binary by the anomalous MB mechanism of Ap/Bp stars. To account for the high mass-transfer rates of most observed persistent NS LMXBs, \citep{Van19a} found the presence of a scaling of the magnetic field with the convective turnover time, and a scaling of MB with the wind mass-loss rate can reproduce persistent LMXBs well. Subsequently, \cite{Van19b} showed the Convection And Rotation Boosted (CARB) MB prescription can successfully reproduce the observed mass-transfer rates in the orbital period versus mass ratio plane for all of the detected Galactic persistent NS LMXBs. Subsequently, the CARB MB model was proposed to be successful in the formation of binary pulsars \citep{Deng21}.

The CARB MB prescription is very successful in solving the mass-transfer rate problem of LMXBs. Some LMXBs were detected to be experiencing a rapid orbital decay (or expansion) in a long timescale (see also Table 1). However, so far it has still not been tested whether or not the orbital evolution predicted by the CARB model is compatible with the observation. In this work, we attempt to investigate this issue by a detailed binary evolution model. We summarize some LMXBs with orbital period derivatives and discuss their influencing factors in theory in Section 2. Section 3 describes the CARB model and stellar evolution code. Section 4 presents the simulated results. Finally, we give a discussion and make a brief summary in Sections 5 and 6, respectively.

\section{Theory and observations of orbital period changes in LMXBs}
\subsection{Determining factors of orbital period changes}
The orbital angular momentum of a LMXB can be expressed as
\begin{equation}
J=\frac{M_1 M_2}{M_2+M_2} \Omega a^{2}
\end{equation}
where $\Omega$ is the orbital angular velocity of the binary, $a$ is the orbital separation, $M_1$ and $M_2$ represent the masses of the accretor and the donor star, respectively. When the donor star fills its Roche lobe in the X-ray binary stage, the H-rich material on the surface of the donor star is transferred onto the accretor through the inner Lagrangian point. If the mass-growth rate of the accretor is $\dot M_1=-f\dot M_2$, the remaining material in unit time ($-\dot M_1-\dot M_2$) is generally thought to be ejected as an isotropic fast wind in the vicinity of the accretor, carrying away its specific orbital angular momentum. According to Kepler's third law ($\Omega=\sqrt{G(M_1+M_2)/a^{3}}$, here $G$ is the gravitational constant) and equation (1), the orbital-period derivative can be derived to satisfy
\begin{equation}
\frac{\dot P}{P} =3\frac{\dot J}{J}-3\frac{\dot M_{2}}{M_{2}}\left[1-qf-\frac{q(1-f)}{3(1+q)}\right],
\end{equation}
where $q=M_2/M_1 $ is the mass ratio of the binary, $P=2\pi/\Omega$ is the orbital period of the binary. 

Since $\dot{J}<0$, the first term in the right-hand side of equation (2) produces a negative orbital-period derivative. On the contrary, the second term results in a positive orbital-period derivative because of a negative mass-loss rate of $\dot{M}_2$ if $g(q,f)=1-qf-\frac{q(1-f)}{3(1+q)}>0$. In general, the evolutionary tendency of the orbit depends on the competition between the first and second terms in equation (2). As a consequence, the orbital evolution of an LMXB can provide much information about the physics of the stellar components and their interactions. During the evolution of LMXBs, the evolution of $\dot{J}/J$, $\dot{M}_2/M_2$, and $g(q,f)$ are very complicated. To study the evolution of the orbital-period derivative, it is inevitable to perform a detailed binary evolution model for LMXBs.

\subsection{Some LMXBs with detected orbital period derivatives}
\begin{table*}
\begin{center}
\caption{Some observed parameters of nine X-ray binaries 
\label{tl}}
\begin{tabular}{@{}ccccccccccccc@{}}
\hline\hline\noalign{\smallskip}
${\rm Sources}$ & $\rm Type  $ & $M_{1}$ &  $M_{2}$ & $P$ & $\dot P$ & Donor Star& $\rm References $ \\
			
&  &   ($M_{\odot}$)  &  ($M_{\odot}$)     & (day) & ($\rm s\,s^{-1})$&Spectral Type  \\ 
\hline\noalign{\smallskip}
			
XTE J1118+480 & BH & ${7.46} _{-0.69}^{+0.34}$ & $0.18\pm 0.06$ & 0.1699  & $-(6.01\pm1.81) \times 10^{-11} $  & K7V-M1V & 1-5\\

A0620-00 & BH & ${6.61} _{-0.17}^{+0.23}$ & $0.40\pm 0.01$ & 0.3230  & $-(1.90\pm0.26) \times 10^{-11} $  & K5V-K7V & 5-7\\

Nova Muscae 1991 & BH & ${11.0} _{-1.4}^{+2.1}$ & $0.89\pm 0.18$ & 0.4326  & $-(6.56\pm4.03) \times 10^{-10} $  & K0V-K4V & 8-11\\
\hline			
SAX J1808.4-3658& NS& ${1.5} _{-0.3}^{+0.6}$ & ${0.05} _{-0.03}^{+0.05}$   & 0.084  & $-(2.82\pm 0.69) \times 10^{-13} $  &$\cdots$ & 12-15\\

X1658-298& NS & $1.48 \pm 0.22$ & $ 0.9\pm 0.3$   & 0.297  & $-(7.2\pm1.8) \times 10^{-11} $ & later than K2V & 16-19 \\
			
AX J1745.6-2901& NS & 1.4 & 0.8             & 0.348  & $-(4.03\pm0.32) \times 10^{-11} $  &$\cdots$  & 20,21 \\
			
\hline
XTE J1710-281& NS & $1.4 $ & $0.28_{-0.14}^{+0.15}$            & 0.137  & $(0.7\pm2.9) \times 10^{-13} $  &$\cdots$ & 22,23 \\

SAX J1748.9-2021& NS & $1.8 \pm 0.6 $ & $0.76_{-0.06}^{+0.07}$            & 0.365  & $(1.1\pm0.3) \times 10^{-10} $  &$\cdots$ & 24-26 \\

2A 1822-371& NS & ${1.96} _{-0.35}^{+0.36}$ &  $0.5 \pm 0.06 $  & 0.232  & $(1.426\pm0.026 )\times 10^{-10} $ & K0V-M2V & 27,28\\
			
\hline\noalign{\smallskip}
\end{tabular}
\tablenotetext{}\\{References. (1) \cite{Remi00}, (2) \cite{Torr04}, (3) \cite{Gonz12}, (4)\cite{Khar13}, (5) \cite{Gonz14}, (6) \cite{McCl86}, (7) \cite{Cant10}, (8) \cite{Remi92}, (9) \cite{Wu15}, (10) \cite{Wu16},(11) \cite{Gonz17},  (12) \cite{Chak98}, (13) \cite{Illi23}, (14) \cite{Good19}, (15) \cite{Bild01}, (16) \cite{Lewi76}, (17) \cite{Comi84}, (18) \cite{Comi89}, (19) \cite{Wach00}, (20) \cite{Hyod09}, (21) \cite{Pont17},
(22) \cite{Mark98}, (23) \cite{Iari24b}, (24) \cite{Kaar03}, (25) \cite{Sann16}, (26) \cite{Cade17}}, (27) \cite{Muo05}, (28) \cite{Iari24a}
\end{center}
\label{tl}
\end{table*}

Table 1 summarizes the observed parameters of nine LMXBs. Three BH LMXBs including XTE J1118+480 (1118), A0620-00 (0620), and Nova Muscae 1991 (1991) were detected to be experiencing a rapid orbital decay. It is intriguing that the detected orbital period derivative of 1991 is an order of magnitude greater than those of 1118 and 0620.

Three NS LMXBs including SAX J1808.4-3658 (1808),  X1658-298 (1658), and AX J1745.6-2901 (1745) were found to be experiencing an orbital decay. In contrast, the other three NS LMXBs 2A 1822-371 (1822), SAX J1748.9-2021 (1748), and  XTE J1710-281 (1710) appear as a tendency of orbital expansion, while the error bar of orbital period derivative of 1710 is very large. 


\section{Stellar Evolution Code and the CARB MB Model}
\subsection{Stellar Evolution Code}
We utilize a binary update version in the Modules for Experiments in Stellar Astrophysics (MESA) code \citep[version r-12115, ][]{paxt11,paxt13,paxt15,paxt18,paxt19} to model the formation and evolution of LMXBs. In calculations, the accretor (NS or BH) is regarded as a point mass, and the donor star is assumed to be a MS star with a solar composition (i.e. $X = 0.7, Y = 0.28, Z = 0.02$) in a circular orbit. The MESA code continuously models the evolution of LMXBs with different initial accretor masses, donor-star masses, and initial orbital periods until the stellar age is greater than the Hubble time ($1.4\times10^{10}$ years) or the time step reaches a minimum time step limit.

During the evolution of the MS donor star, we adopt the “Dutch” wind setting options with a scaling factor of 1.0 in the schemes including $hot_{-}wind_{-}scheme$, $cool_{-}wind_{-}RGB_{-}scheme$, and $cool_{-}wind_{-}AGB_{-}scheme$ \citep{gleb09}. Once the donor star starts the He burning, Type 2 opacities are used for extra C/O burning.

The accretion efficiency of the accretor is assumed to be $\beta$. 
Furthermore, it is generally thought that the mass-growth rate of the accretor is limited by the Eddington accretion rate $\dot{M}_{\rm Edd}$, hence the mass-growth rate of the accretor is $\dot M_1=\min (-\beta\dot M_2, \dot M_{\rm Edd}) $. Same as section 2.1, the excess material in unit time ($-\dot M_1-\dot M_2$) is assumed to be ejected from the vicinity of the accretor, carrying away its specific orbital angular momentum. In this work, the accretion efficiencies of the NS and BH are assumed to be $\beta=0.5$ \citep{Pods02}, and 1.0, respectively. We adopt the Eddington accretion rate of the NS and BH as $1.5\times10^{-8}~M_\odot\,\rm yr^{-1}$ and $\dot{M}_{\rm Edd}=4\pi GM_1/(\kappa c \eta)$ ($c$ is the speed of light in
vacuum, $\kappa$ is the Thompson-scattering opacity of electrons, and $\eta$ is the energy conversion efficiency of the accreting BH), respectively. 

 During the evolution, the total loss rate of orbital angular momentum of LMXBs is given by $\dot J=\dot J_{\rm gr}+\dot J_{\rm ml}+\dot J_{\rm mb}$, where $ \dot J_{\rm gr}$, $ \dot J_{\rm ml}$, $ \dot J_{\rm mb}$ represent the rates of angular-momentum loss caused by GR, mass loss, and MB, respectively. The CARB MB mechanism is arranged to work once the donor star has both a convective envelope and a radiative core \citep{paxt15}.

\subsection{CARB MB Model}
In the MB model, the stellar winds are thought to be ejected at the Alfv\'{e}n radius ($R_{\rm A} $), in which the magnetic energy density of the donor star is equal to the kinetic energy density of the accreting material. Under an assumption of spherical symmetry, the rate of angular momentum loss is $\dot{J}_{\rm MB}=-2\dot{M}_{\rm w} R_{\rm A}^{2}\Omega/3$, where $\dot{M}_{\rm w}$ is the wind-loss rate of the donor star. The Alfv\'{e}n radius is related to the velocity of the wind material at the surface of the donor star, which is generally thought to be the escape velocity. Considering the rotational effects, the escape velocity $v_{\rm esc}$ was modified to be $(v_{\rm esc}^2+2\Omega^{2}R^{2}/K^{2})^{1/2}$ \citep{Matt12}, where $R$ is the radius of the donor star, $K=0.07$ is a constant derived from a grid of simulations \citep{Revi15}. Meanwhile, a convective turnover scaling relation for the magnetic field of the donor star was also included \citep[see also ][]{Van19a,Van19b}. Therefore, the modified MB prescription was named Convection And Rotation Boosted (CARB) MB model \citep{Van19b}.

Including the influence of the donor star's rotation on
the wind's velocity, and the effect of the donor star’s rotation and convective eddy turnover timescale on the surface magnetic field, the rate of angular momentum loss is given by \citep{Van19b}
\[
\dot J_{\rm MB}=-\frac{2}{3} \dot M_{\rm w}^{-1/3} R^{14/3} (v_{\rm esc}^{2}+2\Omega^{2} R^{2}/K^{2})^{-2/3} \]
\begin{equation}
\times  \Omega_{\odot} B_{\odot}^{8/3}(\frac{\Omega}{\Omega_{\odot}}) ^{11/3}(\frac{\tau _{\rm conv}}{\tau _{\odot,\rm conv}} )^{8/3},
\end{equation}
where $\tau_{\rm conv}$ is the turnover time of convective eddies of the donor star, $\Omega_{\odot}\approx 3\times 10^{6}~\rm s^{-1} $, $B_{\odot}=1.0~\rm G$, and $\tau _{\odot,\rm conv}=2.8\times10^{6} ~\rm s$ are the surface rotation rate, the surface magnetic field strength, and the convective-turnover time of the Sun, respectively.

\section{Simulated Results}
To perform the best model that can reproduce the observed properties (donor-star mass, orbital period, and orbital-period derivative) of nine LMXBs, we use the MESA code to simulate a grid of binary evolution for an initial donor-star mass range from 1.0 to $3.0~M_\odot$ (in steps of $0.1~M_\odot$) and an initial orbital period range from 0.6 to 5.0 days (in steps of $0.2~\rm days$). Once the simulated results are approximately consistent with the observed parameters, we then fine tune the initial orbital period or initial donor-star mass.

Figure \ref{f1.eps} summarizes the evolutionary tracks that could match the observed parameters of six known NS LMXBs. The detected donor-star masses and orbital periods of six sources are consistent with our simulated evolutionary tracks at different mass-transfer timescales ($t-t_{\rm RLOF}=9.0\times 10^{6}, 3.9\times 10^{7}, 7.0\times 10^{7}, 8.7\times 10^{7}, 1.3\times 10^{8}, \rm{and}~9.0\times 10^{8} ~ \rm yr$ for 1822, 1808, 1658, 1748, 1710, 1745, respectively). The observed period derivatives of three sources with negative period derivatives are also in good agreement with the calculated results. However, 1822 is the unique source whose $\dot{P}$ can be accounted for by the CARB MB model in three sources with positive period derivatives. In our simulation, 1748 emerges in the orbital shrinkage stage, which is in contradiction with the observed $\dot{P}$. It seems that the period derivative of 1710 is approximately consistent with the simulation due to the visual effect in the logarithmic coordinate. Actually, our simulated period derivative is much higher than the observed value at the current orbital period of 1710 (see also Figure \ref{NSPD}). At the current state, the donor stars in six NS LMXBs are transferring the surface material onto the NSs at a rate of $\sim10^{-8}-10^{-7}~M_\odot\,\rm yr^{-1}$, which is an order of magnitude higher than the one given by the standard MB model \citep{Pods02,shao15}.

Same as Figure \ref{f1.eps}, Figure \ref{f2.eps} compares our best models with the observations of three known BH LMXBs. At the mass-transfer timescale $t-t_{\rm RLOF}=4.3\times 10^{7}, 5.2\times 10^{7},  \rm {and}~5.6\times 10^{7} \ \rm yr$, our simulated BH-MS binaries could evolve into BH LMXBs whose masses and orbital periods are in line with the observed parameters of 1991, 1118, and 0620, respectively. At the current state, the orbital periods of those systems are decreasing,  which is consistent with the observed negative period derivative. Our simulated period derivative is in good agreement with the observed one of 1118, while the calculated value is higher than the detected period derivative of 0620. For 1991, the observed period derivative is approximately two orders of magnitude higher than the simulated value, implying that there exists an efficient loss mechanism of angular momentum. At the current state, the mass-transfer rates of three BH LMXBs are $\sim10^{-9}-10^{-8}~M_\odot\,\rm yr^{-1}$, which are an order of magnitude lower than those in six NS LMXBs.

\begin{figure}[htbp]
\centering
\subfigure{
\includegraphics[width=1.05\linewidth]{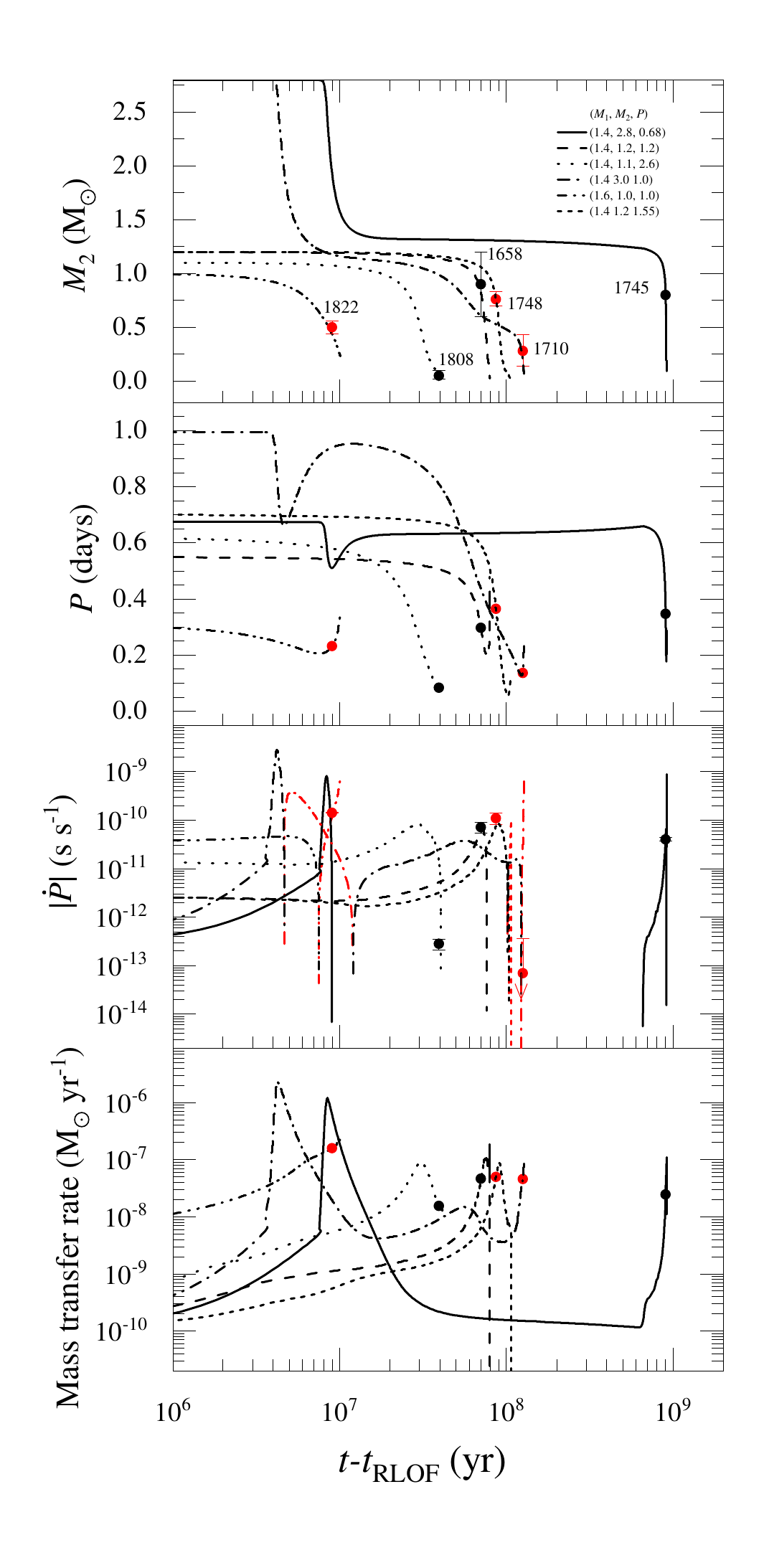} 
}
\label{f1.eps} 

\caption{Evolution of the donor-star mass ($M_2$), orbital period ($P$), orbital period derivative ($\dot{P}$), and mass-transfer rate ($ \dot{M}_2$) with the mass-transfer timescale ($t-t_{\rm RLOF}$) of our simulated six NS X-ray binaries. $t$ and $t_{\rm RLOF}$ are the stellar age and the stellar age when the RLOF occurs. The black and red solid circles with error bars denote those known NS LMXBs with a negative and a positive period derivative, respectively. In the panel of $\dot{P}$, the black and red curves correspond to the evolutionary stage with $\dot{P}<0$ and $\dot{P}>0$, respectively, and we only plot the evolutionary stages with $\dot{P}<0$ of three NS LMXBs which can match the observation of three known sources with negative $\dot{P}$ for clarity.} 
\label{f1.eps}
\end{figure}

\begin{figure}[htbp]
\centering
\subfigure{
\includegraphics[width=1.05\linewidth]{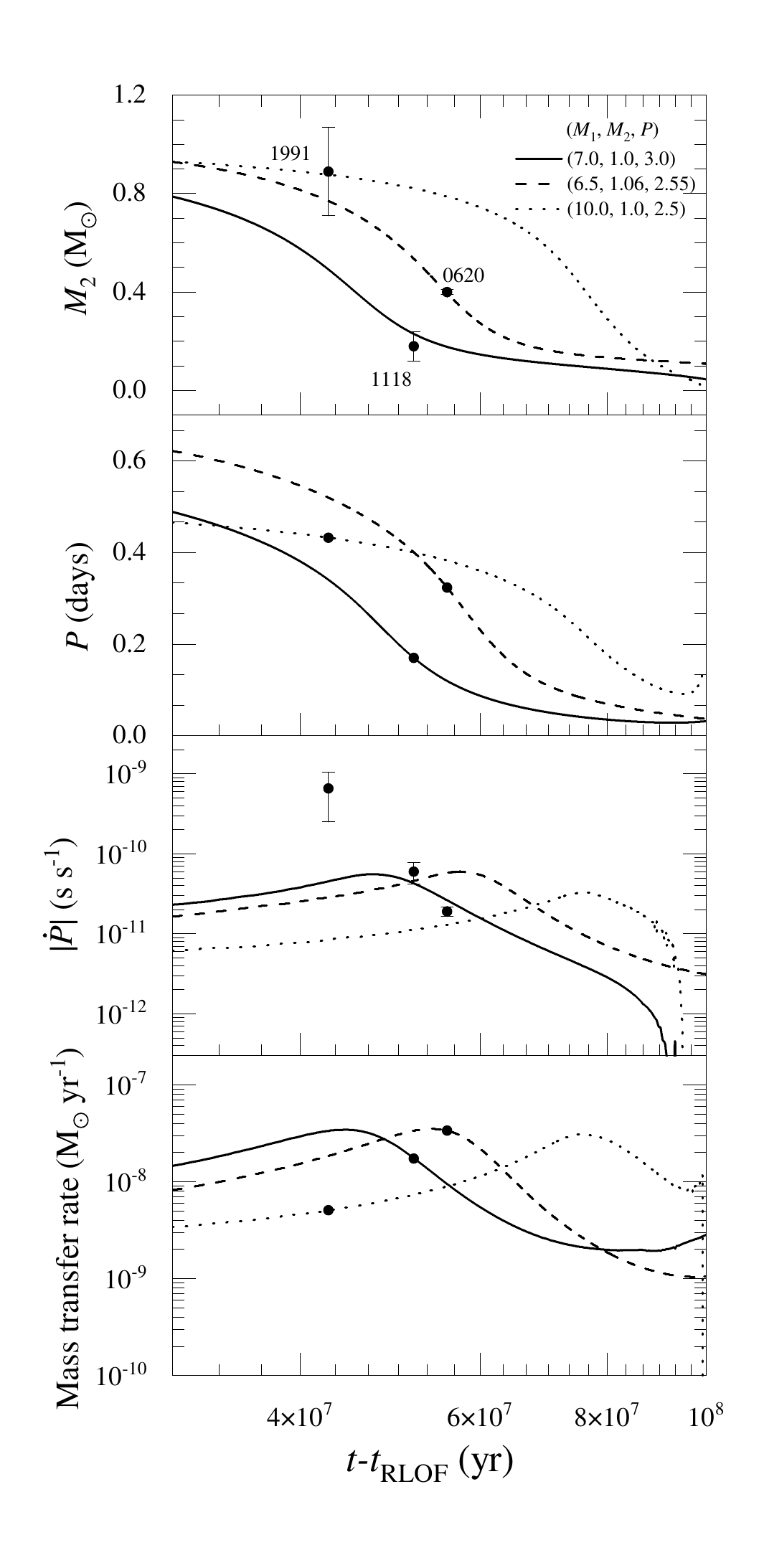} 
}
\label{f2.eps} 

\caption{Same as Figure \ref{f1.eps}, but for our simulated three BH LMXBs. The black solid circles with error bars denote the three known BH LMXBs.} 
\label{f2.eps}
\end{figure}

\begin{figure}
\centering
\includegraphics[width=1.15\linewidth,trim={0 0 0 0},clip]{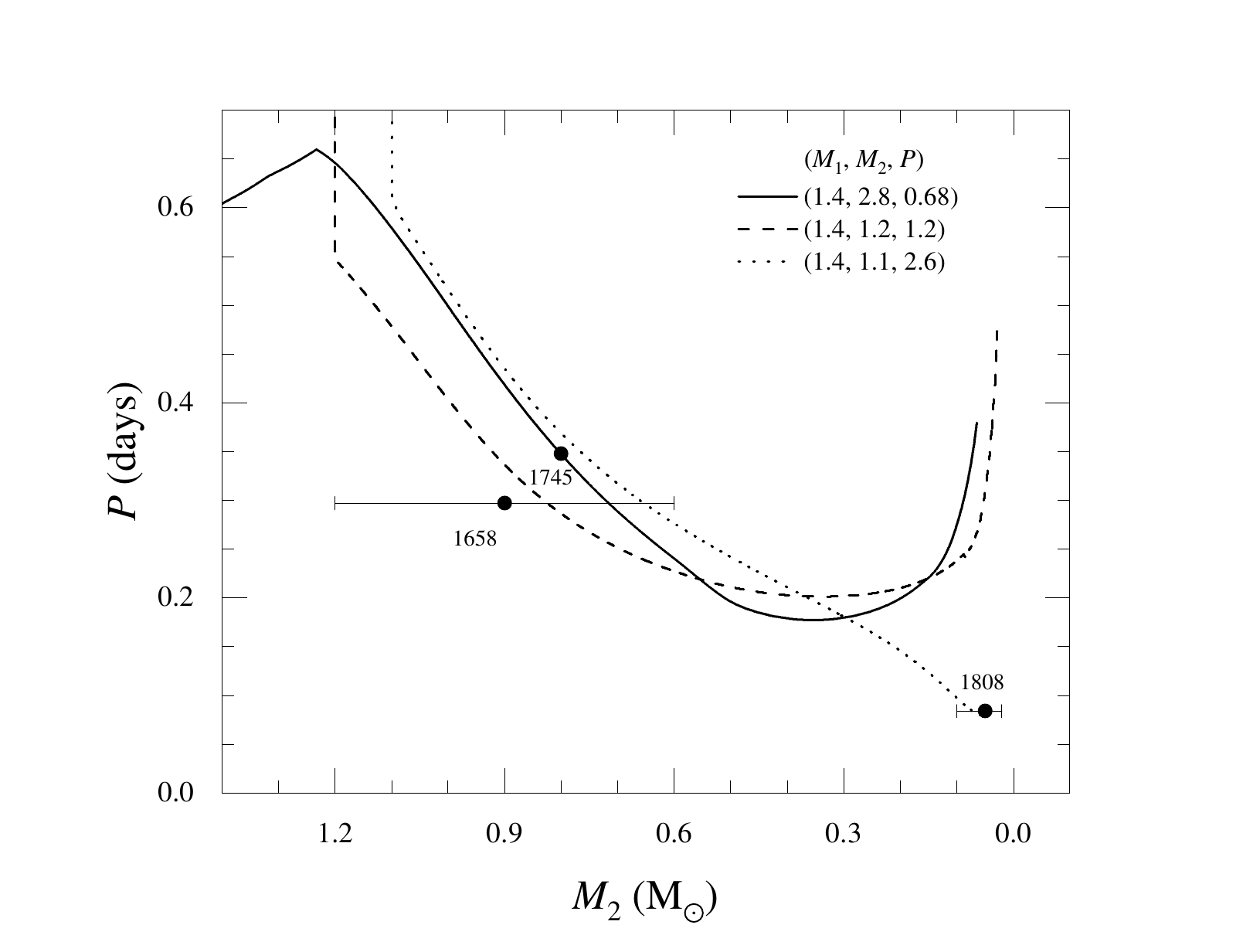}
\includegraphics[width=1.15\linewidth,trim={0 0 0 0},clip]{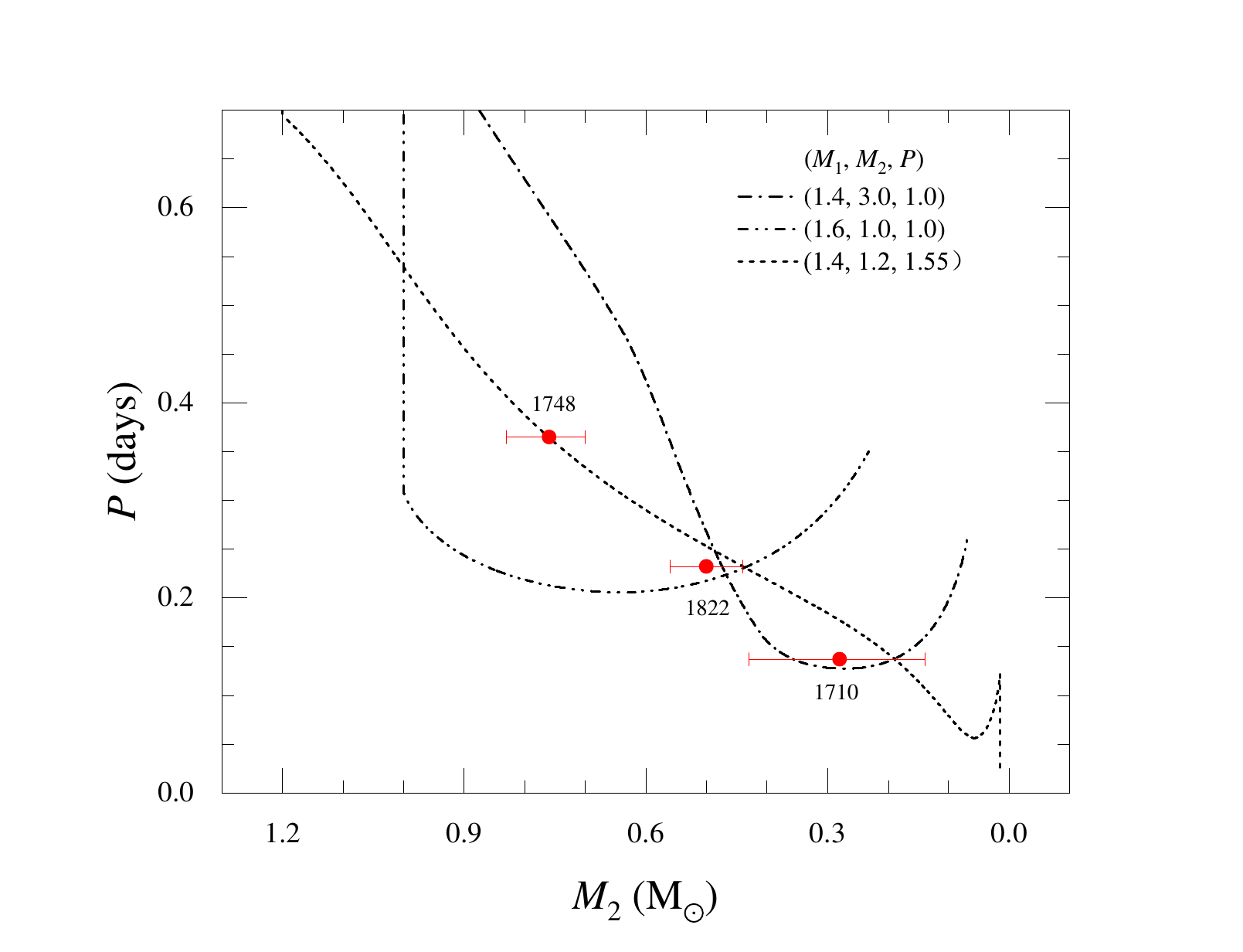}
\caption{Evolutionary tracks of six NS X-ray binaries in the orbital-period vs. donor-star mass diagram.  The black and red solid circles with error bars depict those known NS LMXBs whose orbital period derivatives are negative and positive, respectively. }
\label{NSP}
\end{figure}


\begin{figure}
\centering
\includegraphics[width=1.15\linewidth,trim={0 0 0 0},clip]{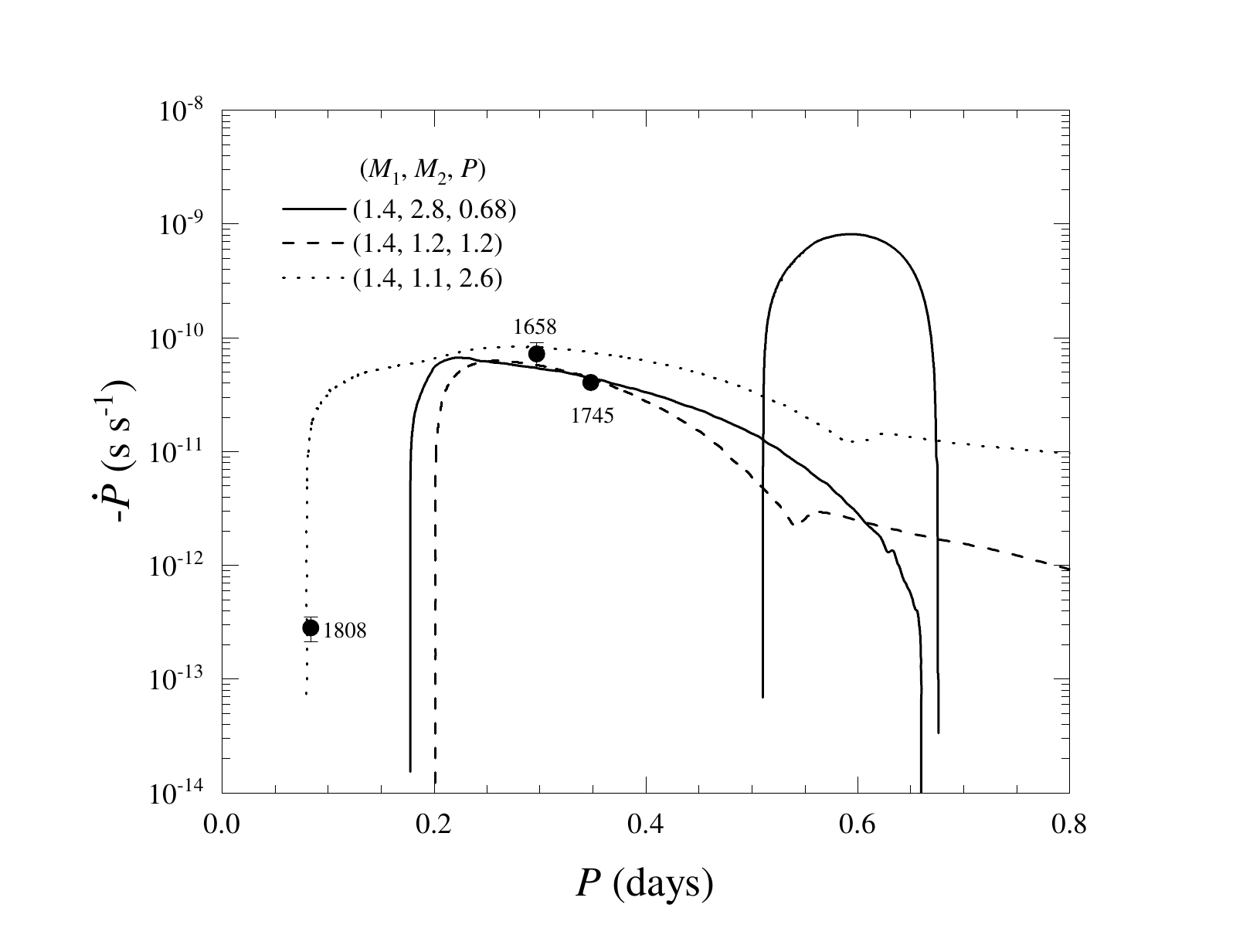}
\includegraphics[width=1.15\linewidth,trim={0 0 0 0},clip]{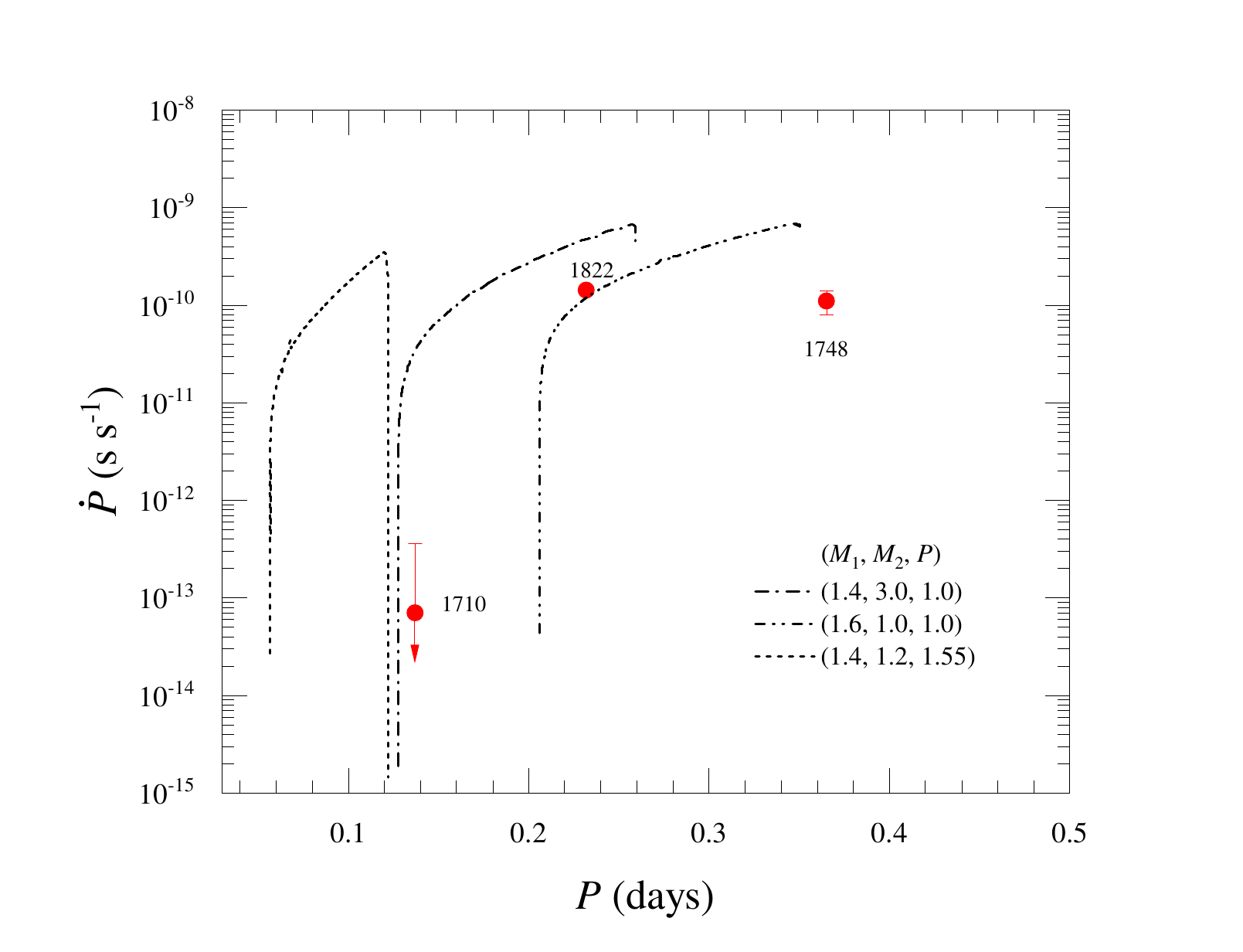}
\caption{Same as in Figure \ref{NSP}, but for the orbital-period derivative vs. orbital period diagram. For clarity, we only plot the evolutionary tracks whose $\dot{P}$ are negative and positive in the top and bottom panels, respectively. } 
\label{NSPD}
\end{figure}

\begin{figure}
\centering
\includegraphics[width=1.15\linewidth,trim={0 0 0 0},clip]{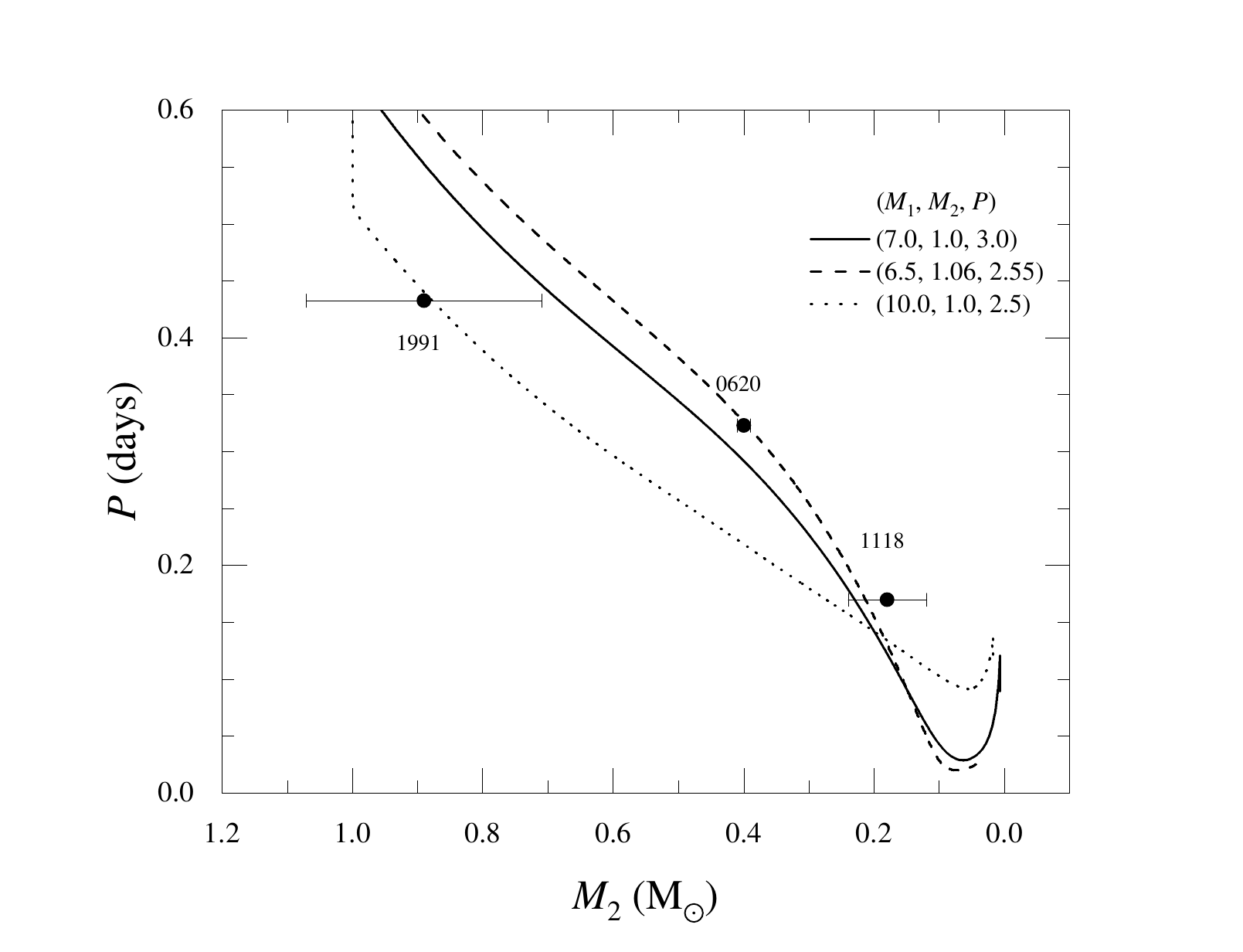}
\includegraphics[width=1.15\linewidth,trim={0 0 0 0},clip]{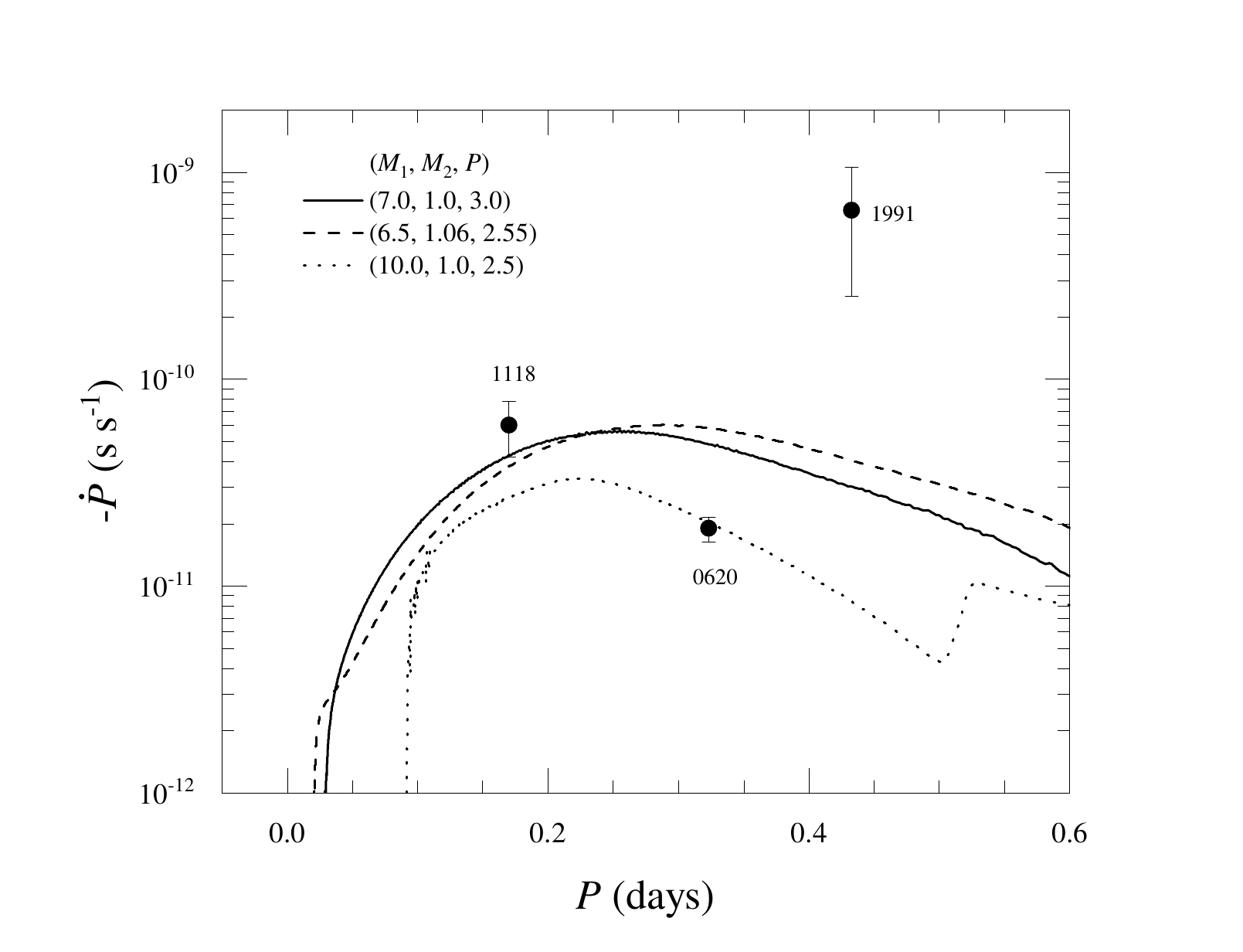}
\caption{Evolutionary tracks of our simulated three BH X-ray binaries in the orbital period vs. donor-star mass diagram (top panel) and the orbital-period derivative vs. orbital period diagram (bottom panel). The solid circles with error bars denote the three known BH LMXBs.} 
\label{BH}
\end{figure}

Figure \ref{NSP} plots the evolutionary tracks of six NS X-ray binaries in the donor-star mass versus the orbital period diagram. In the top panel, three evolutionary tracks with ($M_{1,\rm i}/M_{\odot}$, $M_{2,\rm i}/M_{\odot}$, $P_{\rm i} /\rm days$)= (1.4, 2.8, 0.68), (1.4, 1.2, 1.2), and (1.4, 1.1, 2.6) are in good agreement with the observed parameters of three sources (with negative period derivatives) including 1745, 1658, and 1808, respectively. In the bottom panel, the current state of three NS LMXBs 1710, 1748, and 1822 (with positive period derivatives) is consistent with the evolutionary tracks with ($M_{1,\rm i}/M_{\odot}$, $M_{2,\rm i}/M_{\odot}$, $P_{\rm i} /\rm days$)=(1.4, 3.0, 1.0), (1.4, 1.2, 1.55), and (1.6, 1.0, 1.0), respectively. However, the source 1748 is in the orbital expansion stage in the observations, which is in contradiction to our modeled long-term orbital shrinkage. The source 1822 is a peculiar NS LMXB, in which the NS was proposed to be born massive \citep[$1.6~M_{\odot}$,][]{wei23}. In our simulation, the progenitor of 1822 is an NS-MS binary with ($M_{1,\rm i}/M_{\odot}$, $M_{2,\rm i}/M_{\odot}$, $P_{\rm i} /\rm days$)=(1.6, 1.0, 1.0), in which the initial orbital period is slightly longer than that ($0.35$ days) in \cite{wei23}. There exists a rapid orbital shortening stage (see also dashed, dotted, and dashed-dotted-dotted curves) before those donor stars fill their Roche lobes, which originates from an efficient angular momentum loss due to the CARB MB. Once the orbital expansion effect caused by the mass transfer conquers the orbital shrinkage effect due to the loss of angular momentum, the orbital periods would gradually lengthen. When the donor stars become fully convective ($M_{2} \sim 0.1-0.3\ M_{\odot} $), the MB stops work. During the orbital shrinkage, the minimum orbital period of our simulated NS LMXBs is $\sim 1.2~\rm hours~(\sim 0.05~\rm days)$, and a higher initial donor-star mass or a longer initial period tends to lead to a short minimum period. In principle, the LMXB with the shortest minimum period evolves from the systems with an initial period slightly smaller than the so-called bifurcation period \citep{Chen20}. The physical reason causing this tendency is as follows: in a pre-LMXB with a long initial orbital period, the core of the donor star would possess a high He abundance due to the nucleosynthesis in a long evolutionary timescale, naturally resulting in a more compact donor star and a correspondingly shorter orbital period \citep{tutu87,lin11}. Similarly, a high-mass donor star tends to form a more compact donor star, resulting in a short minimum period.

\begin{figure}
\centering
\includegraphics[width=1.15\linewidth,trim={0 0 0 0},clip]{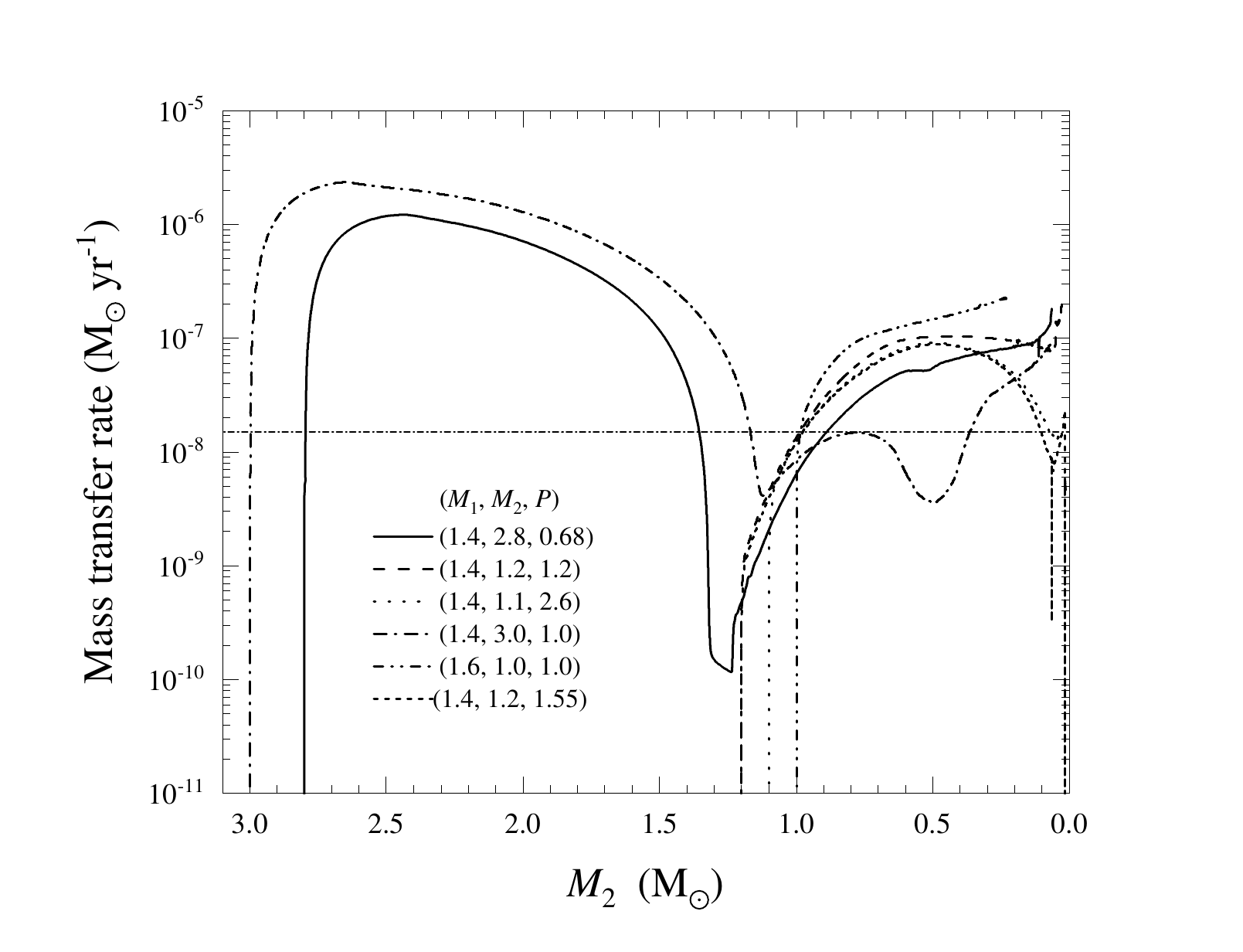}
\includegraphics[width=1.15\linewidth,trim={0 0 0 0},clip]{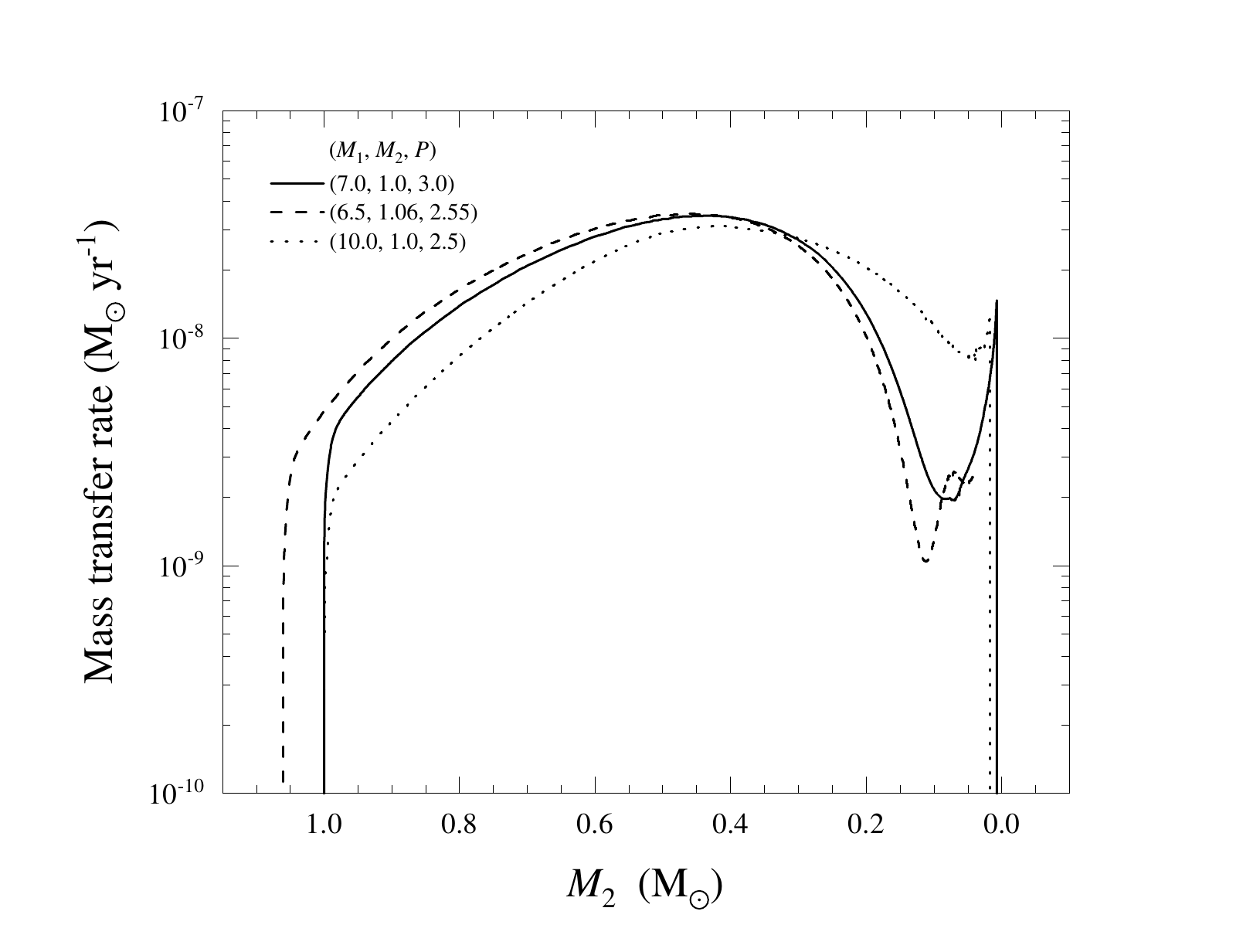}
\caption{Evolution of six NS X-ray binaries (top panel) and three BH X-ray binaries (bottom panel) in the mass-transfer rate vs. donor-star mass diagram. The initial parameters of those X-ray binaries are same as Figures 1 and 2. The horizontal short dotted-dashed line denotes the Eddington accretion rate of an NS.} 
\label{MD}
\end{figure}

Figure \ref{NSPD} displays the evolution of six NS X-ray binaries of Figure \ref{f1.eps} in $\dot P-P$ diagram. In the top panel, the solid, dashed, and dotted curves are in good agreement with the observed orbital-period derivatives and orbital periods of 1745, 1658, and 1808, respectively. The rapid decline of $-\dot{P}$ corresponds to the shortest period, in which three systems evolve to the minimum period of 0.18, 0.2, and 0.08 days for the solid, dashed, and dotted curves, respectively. In the bottom panel, the observed properties of 1822 can be matched by the dashed-dotted-dotted curves, while our simulated period derivative is much higher than the observed one at the current orbital period of 1710. Since the source 1748 can not be interpreted in an orbital expansion stage of $P-M_2$ diagram, the orbital period derivative produced by the CARB MB model is also impossible to fit its observation. Similar to the top panel, the $\dot{P}$ rapidly increases after the orbital period deviates from the minimum period. For the short-dashed curve, a rapid decay of $\dot{P}$ corresponds to the longest period in the orbital expansion stage, after which the binary becomes a detached system. It is determined by the competition between the orbital expansion caused by the mass transfer and the orbital shrinking resulted from the angular momentum loss whether the signs of the orbital period derivative are positive or negative. Because of a high mass-transfer rate ($\sim 10^{-7}~M_{\odot}\,\rm yr^{-1}$, see also Figure \ref{MD}) in the current stage, the contribution of the mass transfer on the orbital evolution exceeds the angular momentum loss, thus the orbital period of 1822 is increasing.

Figure \ref{BH} illustrates the evolutionary tracks of three BH LMXBs in the orbital period versus donor-star mass and the orbital period derivative versus orbital period diagrams. Because 1118 has a slightly heavy BH, the observed parameters of 1118 are consistent with the evolutionary track of a BH LMXB with $M_{1,\rm i}=7.0~M_{\odot}$, $M_{2,\rm i}=1.0~M_{\odot}$, and $P_{\rm i}=3.0~\rm days$. In the $P-M_2$ diagram, the observed values of 0620 can match the evolution of a BH LMXB with $M_{1,\rm i}=6.5~M_{\odot}$, $M_{2,\rm i}=1.06~M_{\odot}$, and $P_{\rm i}=2.55~\rm days$, while our simulated orbital period derivative is much greater than the observed one. In our calculations, the rate of angular momentum loss caused by the CARB MB mechanism is too small to produce the detected orbital-period derivative of 1991. Comparing with those NS LMXBs, BH LMXBs possess a relatively low mass-transfer rate ($\le4.0\times 10^{-8}~M_{\odot}\,\rm yr^{-1}$, see also Figure \ref{MD}), which can only produce a weak orbital expansion effect. Consequently, the orbits of three BH LMXBs continuously shrink until the orbital expansion effect caused by the mass transfer is stronger than the orbital shrinkage effect due to the gravitational radiation. The minimum period of BH LMXBs is $\sim 0.5~\rm hours~(\sim 0.025~\rm days)$, which is shorter than the one of our simulated NS LMXBs. This discrepancy originates from the different initial orbital periods between NS and BH LMXBs, in which the initial periods of three BH LMXBs are longer than those of NS LMXBs. As mentioned above, a relatively long initial period tends to result in a short minimum period if the initial period is less than the bifurcation period.

\begin{figure}
\centering
\includegraphics[width=1.15\linewidth,trim={0 0 0 0},clip]{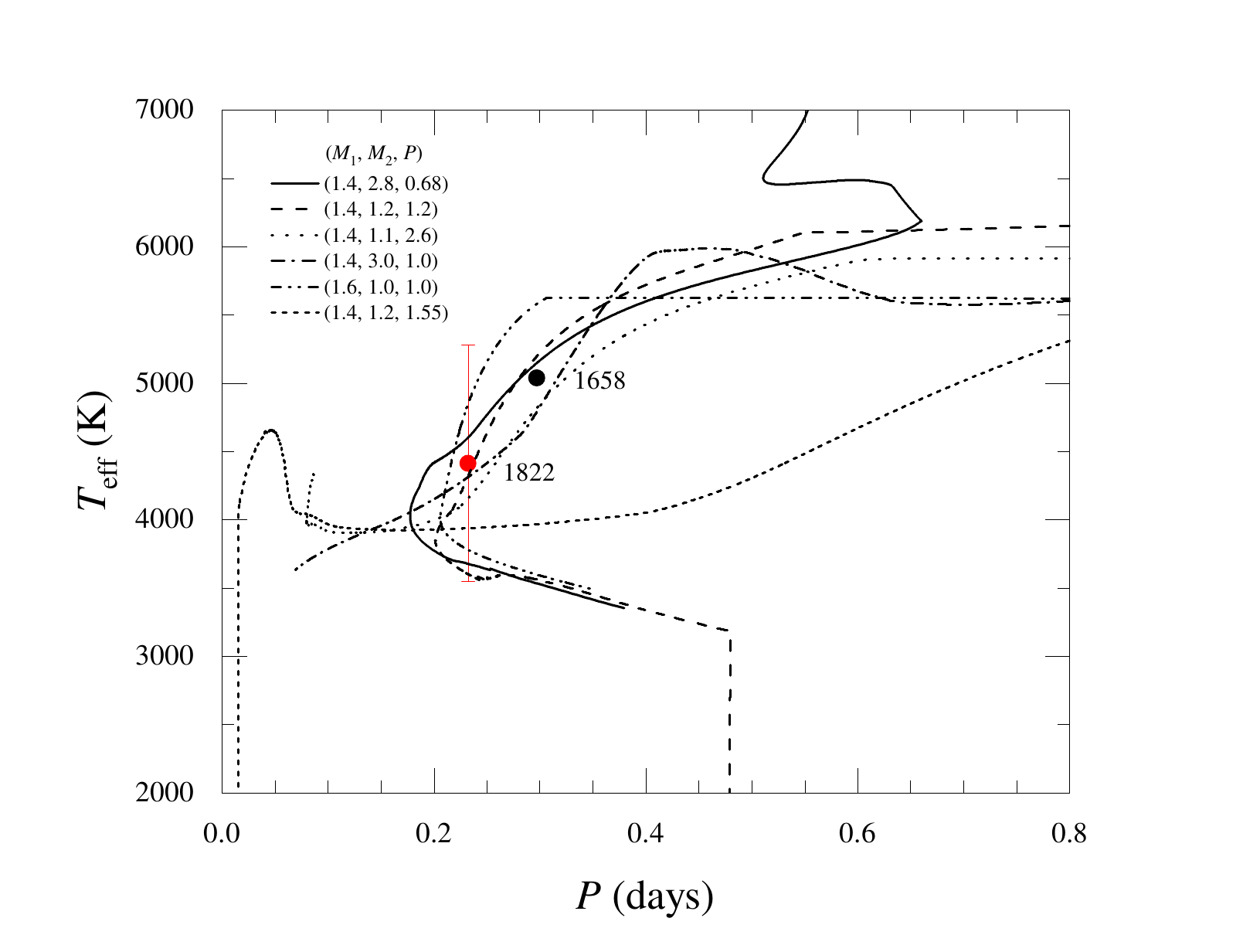}
\includegraphics[width=1.15\linewidth,trim={0 0 0 0},clip]{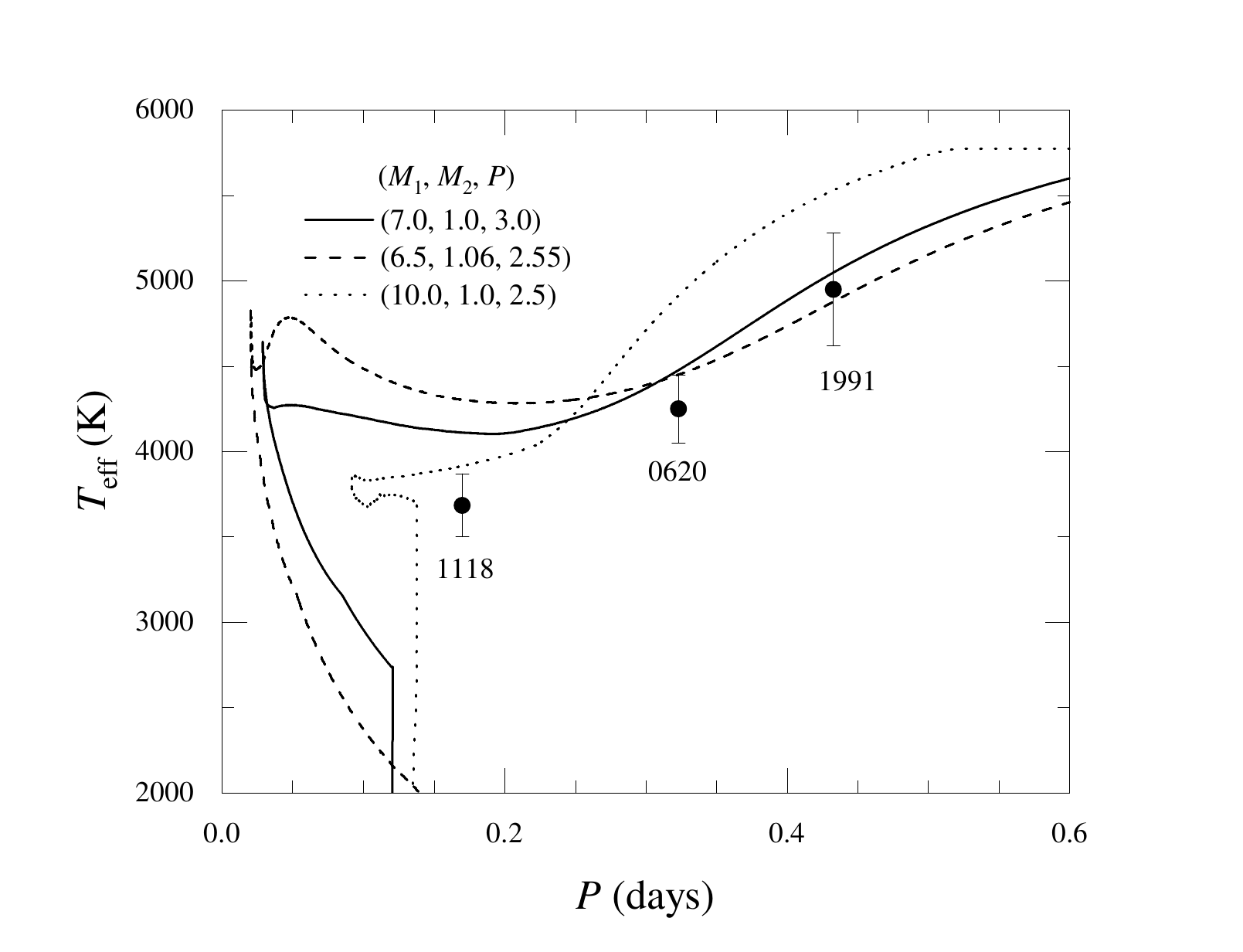}
\caption{Evolution of NS X-ray binaries (top panel) and BH X-ray binaries (bottom panel) in the $T_{\rm eff}-P$ diagram. The solid squares and solid circles with error bars denote the two NS LMXBs and three BH LMXBs with a known spectral type. } \label{PT}
\end{figure}

The evolution of the mass-transfer rates of NS and BH X-ray binaries is shown in Figure \ref{MD}. In the two intermediate-mass X-ray binaries with $M_{2,\rm i}=2.8$ and $3.0~M_{\odot}$, the mass transfer first proceeds on the thermal timescale of the donor star at a rate of $\sim 10^{-7}-10^{-6}~M_{\odot}\,\rm yr^{-1}$ because the material is transferred from the more massive donor star to the less massive NS. During the mass ratio near to 1, the mass transfer rates gradually decrease. Since the system with $M_{2,\rm i}=3.0~M_{\odot}$ has a slightly long initial orbital period, the long-term nuclear evolution drives the MS donor star to a deeply evolved stage as $q\sim 1$, in which the donor star transfers the material onto the NS at a rate much higher than the system with $M_{2,\rm i}=2.8~M_{\odot}$. Subsequently, the MB mechanisms start work because the donor stars develop a convective envelope when their masses are $\sim 1.5~M_\odot$, and two systems initiate the second mass transfer stage. Finally, the mass transfer rates sharply decline because the MB mechanism stops due to the vanishing of radiative core the donor stars become completely convective). Because of an initial mass ratio less than 1, the evolution of the mass-transfer rates in other four NS X-ray binaries with low-mass donor stars and three BH X-ray binaries is similar to the second mass transfer stage of two NS X-ray binaries with massive donor stars. When $M_2=1.1~M_\odot$, the MB drives two systems with ($M_{1,\rm i}/M_{\odot}$, $M_{2,\rm i}/M_{\odot}$, $P_{\rm i} /\rm days$)=(1.4, 1.1, 2.6) and (1.4, 1.2, 1.55) evolve to a similar orbital period ($\sim 0.6~\rm days$, see also Figure \ref{NSP}), causing almost overlapped evolutionary tracks (see also the short-dashed and dotted curves). In some systems, the donor stars eventually decouple from their Roche lobe after the H-rich envelope is fully stripped. As a consequence, the mass-transfer rates sharply decrease and the mass transfer ceases. To fit the observed data, our simulate system with $M_{1,\rm i}=7.0~M_{\odot}$, $M_{2,\rm i}=1.0~M_{\odot}$, and $P_{\rm i}=3.0~\rm days$ has a mass-transfer rate of $\sim 1.0\times10^{-8}~M_{\odot}\,\rm yr^{-1}$ in the current stage of 1118, which is approximately a order of magnitude higher than that ($\sim 2.0\times10^{-9}~M_{\odot}\,\rm yr^{-1}$) in the circumbinary disk model \citep{chen19}. The observed peak X-ray luminosity ($\sim 10^{-3}L_{\rm Edd}$, $L_{\rm Edd}$ is the Eddington luminosity) of 1118 revealed an accretion rate of $\sim 10^{-10}~M_{\odot}\,\rm yr^{-1}$ \citep{wu10}. This implies that the radiation efficiency of the accretion disk in 1118 is extremely low under an advection-dominated accretion flow. Unfortunately, the calculated mass-transfer rate is much higher than the critical rate ($\sim10^{-9}~M_{\odot}\,\rm yr^{-1}$)
for the advection-dominated accretion flow proposed by \cite{nara95}.

Figure \ref{PT} presents the evolutionary tracks of our simulated NS and BH X-ray binaries in the effective temperatures of the donor stars versus orbital periods diagram. The effective temperatures of three BH LMXBs are derived according to the detected spectral types of their donor stars. Our models can only match the observed effective temperature of the donor star in 0620 at the current orbital period. However, the donor stars of 1118 and 1991 are observed to be slightly cooler than our calculated values, which is similar to the previous works \citep[see also][]{Just06,chen06,chen19}. Among NS LMXBs, only the donor stars of two sources (1822 and 1658) have been detected spectral types. Adopting the CARB MB model, our simulated evolutionary tracks with the dashed and dashed-dotted-dotted curves are approximately consistent with the observed data of two NS LMXBs 1658 and 1822, respectively. We also exhibit the evolution of the effective temperatures of the donor stars in other NS LMXBs, which can be compared with and testified by future optical observations.

\begin{figure}
\centering
\includegraphics[width=1.05\linewidth,trim={0 0 0 0},clip]{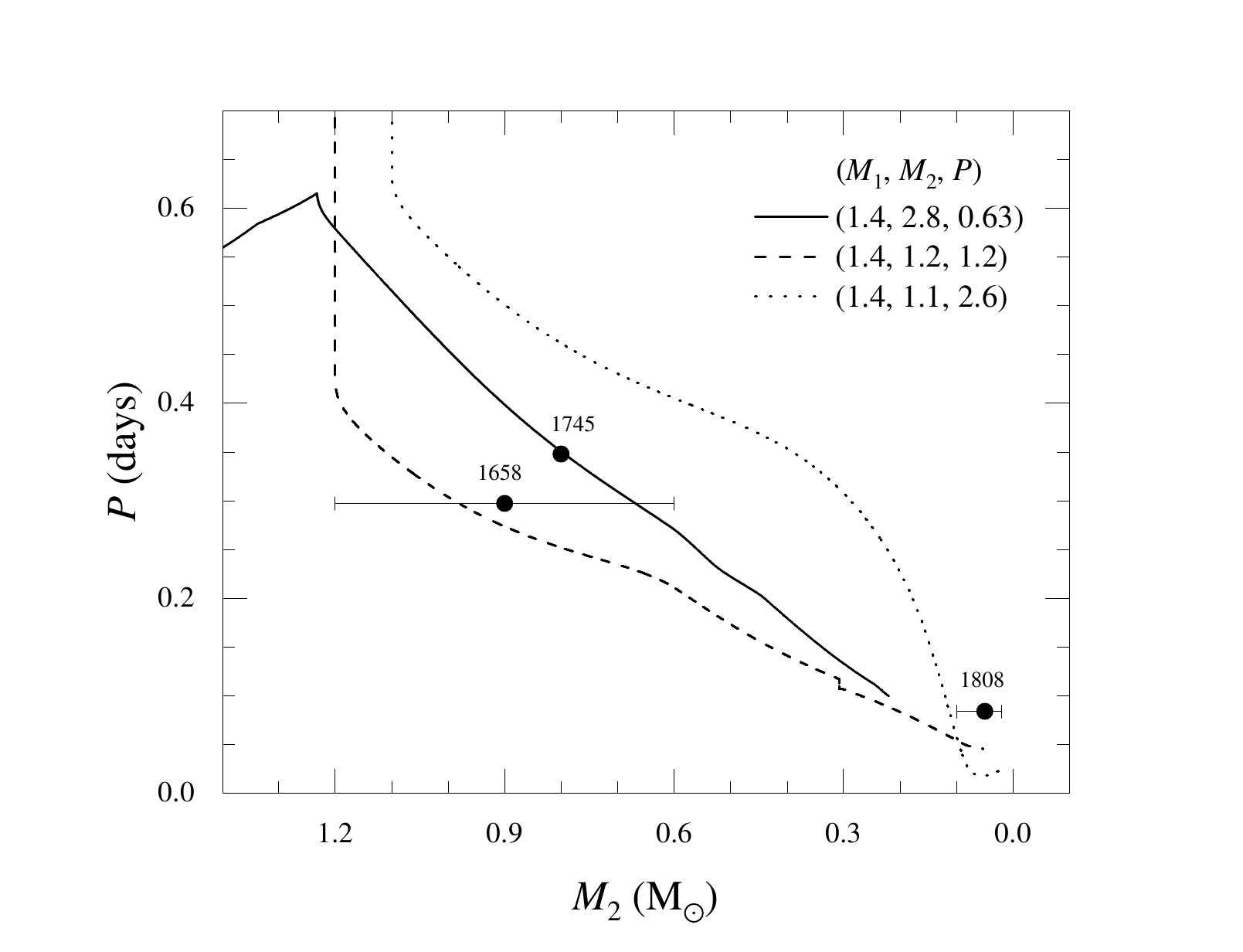}
\includegraphics[width=1.05\linewidth,trim={0 0 0 0},clip]{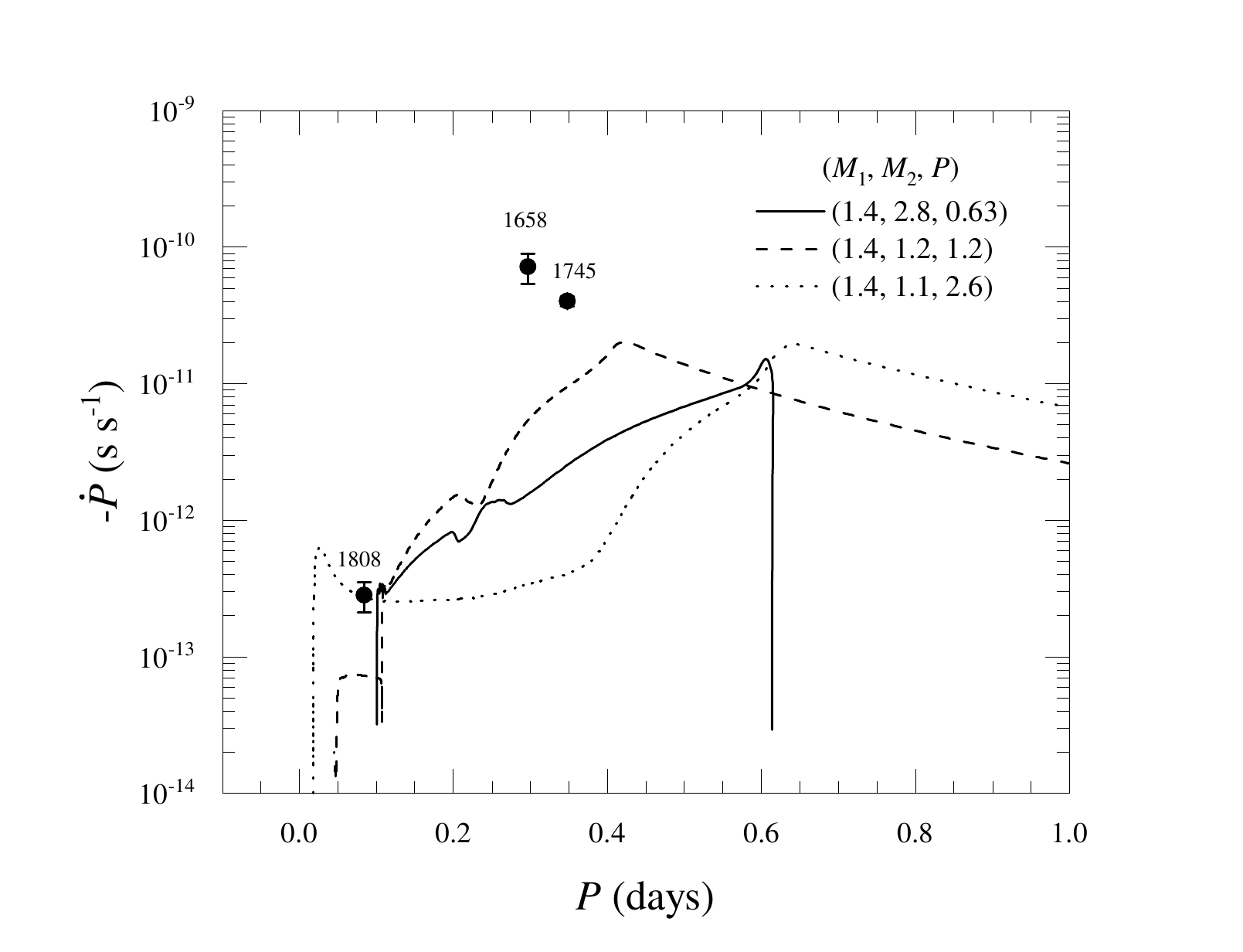}
\includegraphics[width=1.05\linewidth,trim={0 0 0 0},clip]{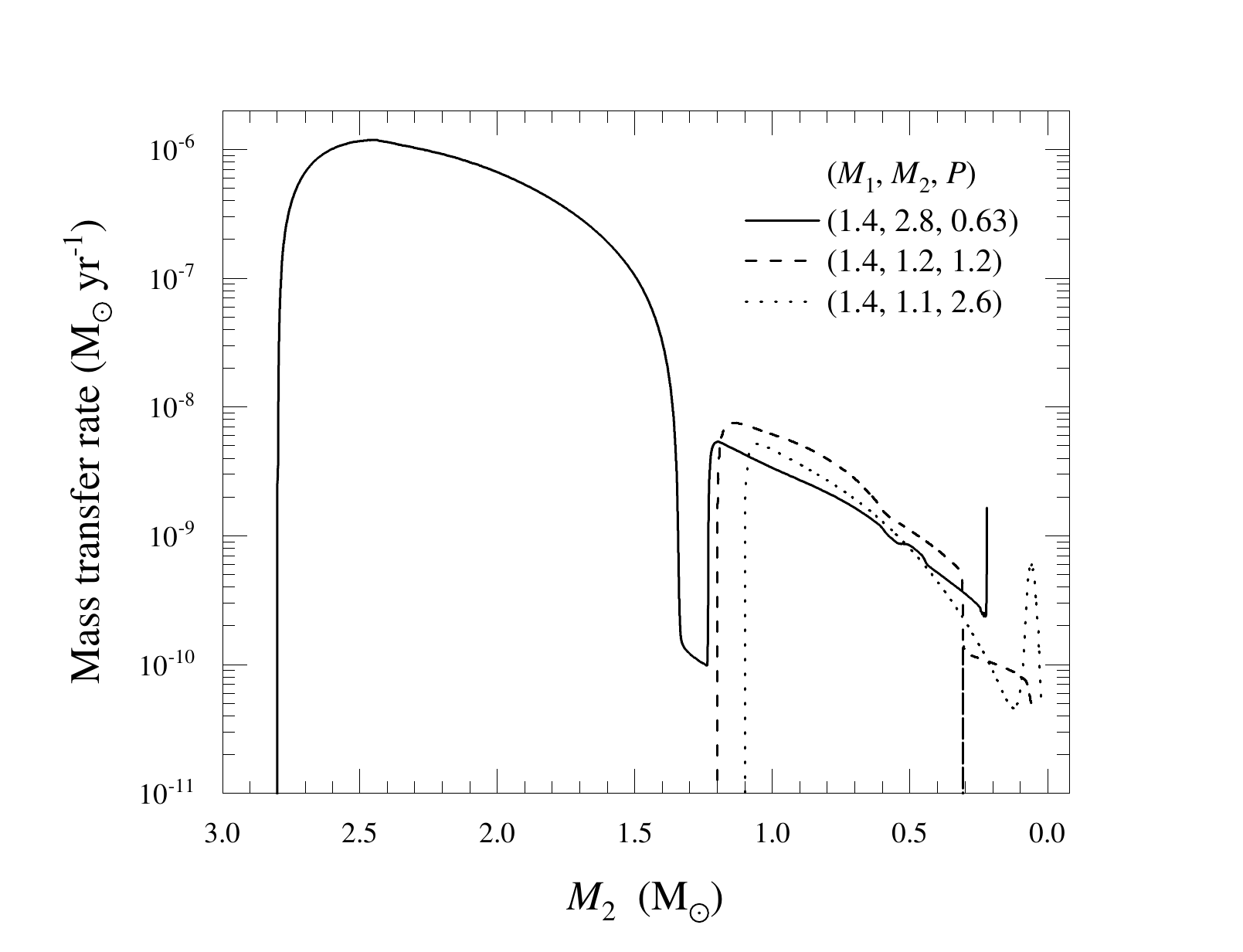}
\caption{Evolution of three NS X-ray binaries in $P-M_2$ (top panel), $-\dot{P}-P$ (middle panel), and $\dot{M}_2-M_2$ (bottom panel) diagrams. The black solid circles with error bars denote the three known NS LMXBs with a negative $\dot{P}$. In this figure, we adopt the standard MB model with $\gamma=4$.} 
\label{f8}
\end{figure}

Actually, the standard MB model could also produce the detected donor-star masses and orbital periods of some LMXBs. In Figure \ref{f8}, we illustrate the evolution of three NS-MS star binaries that can account for the donor-star masses and orbital periods of three known NS LMXBs with a negative $\dot{P}$. In the simulations, the standard MB model given by \cite{rapp83} with $\gamma=4$ is adopted. Taking similar initial parameters of Figure 1, the observed data are approximately consistent with the calculated results in the $P-M_2$ diagram (there exists a tiny difference between the calculated and observed donor-star mass for 1808). However, only the observation of 1808 with a small $\dot{P}$ can match the evolutionary curve in the $\dot{P}-P$ diagram. Two sources 1658 and 1745 were detected relatively high period derivatives, which are approximately $1-2$ orders of magnitude higher than the calculated values at the current period. In the bottom panel, the mass-transfer rates produced by the standard MB model are $\sim 10^{-10}-10^{-9}~M_{\odot}\,\rm yr^{-1}$ when the donor-star masses are in the range of $0.05-1.2~M_\odot$, and are about $1-2 $ orders of magnitude lower than those given by the CARB MB model. This difference is caused by the different loss rates of angular momentum in the two MB models. The CARB MB model has a loss rate of angular momentum much higher than that of the standard MB model. Because the standard MB model can only produce a low mass-transfer rate, it leads to a weak orbital expansion effect. Therefore, it is difficult to interpret the rapid orbital expansion (with a high positive $\dot{P}$) of NS LMXBs like 1822 \citep{wei23}. Furthermore, the $\dot{P}$ predicted by the standard MB model is an order of magnitude smaller than those observed in three BH LMXBs \citep{chen19}. 

\section{Discussions}
\subsection{Formation of BH LMXBs}
Among BH LMXBs, the orbital decay of 1991 can not be interpreted by the CARB MB model. \cite{chen19} proposed that a heavy circumbinary disk with a mass of $10^{-7}~M_\odot$ surrounded 1991 can produce its observed orbital period derivative, while the simulated effective temperature of the donor star is slightly higher than the observed value. Our detailed stellar evolution models found that the CARB MB prescription can reproduce the observed donor star mass, orbital period, and orbital period derivative of 1118. However, similar to the previous works \citep{Just06,chen06,chen19}, there still exist an effective temperature problem, in which the simulated effective temperature is slightly higher than that inferred from the spectral type of the donor star. As an alternative evolutionary channel, the dynamical friction of dark matter could alleviate the effective temperature problem of those BH LMXBs with an orbital period less than 0.3 days \citep{qin24b}. It is noteworthy that the effective temperature problem is still not completely understood in those BH LMXBs with a relatively long orbital period.

\subsection{Other alternative models}
Actually, the anomalous MB of Ap/Bp stars with an intermediate-mass can also produce an efficient angular momentum loss \citep{Just06}, driving a rapid orbital shrinkage or orbital expansion. \cite{chen17} proposed that the anomalous MB of a $1.6-1.8~M_\odot$ Ap/Bp star can result in the high mass-transfer rate observed in Sco X-1.
However, only a small fraction of $5\%$ Ap/Bp stars in A/B stars possesses anomalously strong surface magnetic fields of $10^2-10^4~\rm G$. Therefore, it seems impossible that the anomalous MB mechanism can be responsible for the orbital period changes of all LMXBs.

In principle, a surrounding circumbinary disk can also extract orbital angular momentum from the orbital motion of LMXBs. For example, \cite{wei23} employed the circumbinary disk model to interpret the observed parameters especially orbital period derivative of 1822. In observations, the Wide-Field Infrared Survey Explorer confirmed that the three black hole LMXBs XTE J1118+480, A0620-00, GRS 1915+105, and NS LMXB 3A 1728-247 are surrounded by circumbinary disks \citep{wang14}. However, the orbital period (1160.8 days) of NS LMXB 3A 1728-247 is much longer than those NS LMXBs in Table 1. Therefore, it is still not identified whether some NS LMXBs with a short orbital period are surrounded by a circumbinary disk. 

The dynamical friction between dark matter and the companion stars in LMXBs can produce an efficient loss mechanism of angular momentum, driving the LMXBs to evolve toward short period systems. The dynamical friction due
to a dark matter density spike around stellar mass BHs was thought to be accountable for the fast orbital decay in two BH LMXBs XTE
J1118+480 and A0620-00 \citep{chan23}. Recently, \cite{qin24b} found that the dynamical friction can help BH binaries with an intermediate-mass
companion star to evolve into BH LMXBs, and alleviate the effective temperature problem of BH LMXBs with a relatively short orbital period. However, it is impossible that there exists a density spike of dark matter surrounding NSs. As a consequence, the dynamical friction of dark matter can not be in charge of the period change of those NS LMXBs.

Applegate mechanism could also give rise to a short-term decrease of orbital period rather than a secular decrease \citep{appl92}.  It is generally thought that the stellar magnetic field originates from magnetic dynamo effect, in which the interplay between differential rotation and cyclonic convection causes a magnetic activity cycle. The magnetic activity results in a variable deformation of the donor star by transferring angular momentum in different convective zones of the star, hence alter the rotational velocity of the star. However, the torque redistributing the angular momentum can only be achieved by a mean subsurface magnetic field of $\sim 10^3~\rm G$ \citep{appl92}. Such a magnetic field is much strong for those donor stars in LMXBs. Therefore, it seems that Applegate mechanism can not account for the period change of all LMXBs even if their observed period changes are a short-term phenomenon.

\section{Summary}
Based on the CARB MB prescription, in this work, we investigate the orbital evolution of some NS LMXBs and BH LMXBs with a detected orbital period derivative. Our detailed binary evolution model can successfully reproduce the observed donor star mass, orbital period, and orbital period derivative of four NS LMXBs including 1745, 1658, 1808, and 1822, as well as the BH LMXB 1118. However, the CARB MB prescription can not account for the orbital evolution of two NS LMXBs (1748 and 1710) with positive orbital period derivatives and two BH LMXBs (0620 and 1991). For five LMXBs with a detected spectral type of the donor star, the modeled effective temperatures based on the CARB MB prescription are in line with the observed values of three sources (except for 1118 and 1991). Among three known NS LMXBs with negative $\dot{P}$, the standard MB model can only produce the observed period derivative of 1808. The previous works indicate that the loss rate of angular momentum in the standard MB model is too small to drive a rapid orbital expansion of 1822 \citep{wei23} and orbital shrinkage of three BH LMXBs \citep{chen19}. Furthermore, the CARB MB prescription is also able to produce the observed X-ray luminosities of all persistent NS LMXBs \citep{Van19b,Van21}. Considering the successful test of the CARB MB prescription on the formation and evolution of most LMXBs, it may be a plausible substitute to replace the standard MB model in the stellar and binary evolution models. To further test the validity of the CARB MB prescription, it is necessary to invoke more LMXB samples with detected orbital period derivatives and spectral types of donor stars.

\begin{acknowledgments} We are extremely grateful to the anonymous referee for constructive comments that improved this manuscript. This work was partly supported by the National Natural Science Foundation of China (under grant No. 12273014), and the Natural Science Foundation (under grant number ZR2021MA013) of the Shandong Province.
\end{acknowledgments}

\bibliography{ref}
\bibliographystyle{aasjournal}
\end{document}